\documentclass[11pt]{article}
\usepackage{spconf,amsmath,graphicx}
\usepackage{cite}
\usepackage{bibspacing}
\usepackage{booktabs}
\usepackage{color}
\usepackage{amsfonts}
\usepackage{multirow}
\usepackage{amssymb}
\usepackage{array}
\usepackage{authblk}
\usepackage{threeparttable}
\newcommand{\PreserveBackslash}[1]{\let\temp=\\#1\let\\=\temp}
\newcolumntype{C}[1]{>{\PreserveBackslash\centering}p{#1}}
\newcolumntype{R}[1]{>{\PreserveBackslash\raggedleft}p{#1}}
\newcolumntype{L}[1]{>{\PreserveBackslash\raggedright}p{#1}}
\usepackage{enumitem}
\usepackage{subfigure}
\usepackage[hang]{footmisc}
\usepackage{graphicx} 
\usepackage[belowskip=-1pt]{caption}
\usepackage[skip=3.0pt]{caption}

\AtBeginDocument{
 \footnotesize\setlength{\footnotemargin}{0pt}\normalsize
 \edef\hangfootparindent{\the\parindent}
}

\usepackage{lipsum} % mock text

\makeatletter
\newcommand\email[2][]%
   {\newaffiltrue\let\AB@blk@and\AB@pand
      \if\relax#1\relax\def\AB@note{\AB@thenote}\else\def\AB@note{\relax}%
        \setcounter{Maxaffil}{0}\fi
      \begingroup
        \let\protect\@unexpandable@protect
        \def\thanks{\protect\thanks}\def\footnote{\protect\footnote}%
        \@temptokena=\expandafter{\AB@authors}%
        {\def\\{\protect\\\protect\Affilfont}\xdef\AB@temp{#2}}%
         \xdef\AB@authors{\the\@temptokena\AB@las\AB@au@str
         \protect\\[\affilsep]\protect\Affilfont\AB@temp}%
         \gdef\AB@las{}\gdef\AB@au@str{}%
        {\def\\{, \ignorespaces}\xdef\AB@temp{#2}}%
        \@temptokena=\expandafter{\AB@affillist}%
        \xdef\AB@affillist{\the\@temptokena \AB@affilsep
          \AB@affilnote{}\protect\Affilfont\AB@temp}%
      \endgroup
       \let\AB@affilsep\AB@affilsepx
}
\makeatother

\title{GMP-TL: GENDER-AUGMENTED MULTI-SCALE PSEUDO-LABEL ENHANCED TRANSFER LEARNING FOR SPEECH EMOTION RECOGNITION}

\author[1*]{Yu Pan}
\author[3*]{Yuguang Yang}
\author[3$\dagger$]{Heng Lu}
\author[2$\dagger$]{Lei Ma}
\author[1]{Jianjun Zhao}

\affil[1]{Kyushu University, Japan \quad $^2$University of Alberta, Canada \quad $^3$Ximalaya Inc., ShangHai, China}

\begin{document}
%\ninept
%
\maketitle

\begin{abstract}
The continuous evolution of pre-trained speech models has greatly advanced Speech Emotion Recognition (SER).
However, current research typically relies on utterance-level emotion labels, inadequately capturing the complexity of emotions within a single utterance. 
In this paper, we introduce GMP-TL, a novel SER framework that employs gender-augmented multi-scale pseudo-label (GMP) based transfer learning to mitigate this gap.
Specifically, GMP-TL initially uses the pre-trained HuBERT, implementing multi-task learning and multi-scale k-means clustering to acquire frame-level GMPs. 
Subsequently, to fully leverage frame-level GMPs and utterance-level emotion labels, a two-stage model fine-tuning approach is presented to further optimize GMP-TL.
Experiments on IEMOCAP show that our GMP-TL attains a WAR of 80.0\% and an UAR of 82.0\%,  achieving superior performance compared to state-of-the-art unimodal SER methods while also yielding comparable results to multimodal SER approaches.
\end{abstract}


\section{Introduction}
\label{sec:intro}%

\noindent
\noindent
As one of the crucial elements in realizing human-computer interaction, speech emotion recognition (SER) aims to categorize emotions conveyed through human speech, which has been extensively applied in diverse practical domains \cite{hci}. 

{
\let\thefootnote\relax
\footnote{$*$ denotes equal contributing.}
\footnote{$\dagger$ denotes the corresponding author.}
\footnote{Work done at Ximalaya Inc.}
}

In recent years, the remarkable advancements in machine and deep learning technologies have significantly accelerated the progress of the field of SER.
By constructing robust deep neural networks such as convolutional neural networks or recurrent neural networks \cite{li2019improved,ghriss2022sentiment,TIMNet} based models, these methods can achieve improved performance compared to traditional SER approaches \cite{1326051,5674019}. 
Nonetheless, due to the factors such as the inherent variability and complexity of speech signals \cite{kim2017towards,pan2023msac,pan2024promptcodec} and the scarcity of available large-scale labeled speech emotion data resulting from difficulties in data collection and annotation \cite{Chen,Gat}, achieving accurate and reliable emotion recognition continues to pose challenges.

To tackle these issues, numerous researchers \cite{Morais,yi2023exploring,pan2023gemo,hu2023joint} have attempted to leverage transfer learning by incorporating large-scale pre-trained speech models such as HuBERT \cite{hubert}, Wav2vec 2.0 \cite{wav2vec2}, and WavLM \cite{wavlm} into speech emotion modeling.
For instance, 
Morais \emph{et al}. \cite{Morais} presented a modular end-to-end Upstream + Downstream SER architecture, which facilitates the integration of the pretrained, fine-tuned, and averaged Upstream models. 
Hu \emph{et al}. \cite{hu2023joint} introduced a joint network combining the pre-trained Wav2Vec 2.0 model and a separate spectrum-based model to achieve SER. 
% Nevertheless, despite achieving commendable results, these methods predominantly depend on utterance-level emotion labels as the training objectives, overlooking the local information within each utterance that could provide valuable complementary details for modeling speech emotion.
Nevertheless, despite achieving commendable results, these methods predominantly depend on utterance-level emotion labels as training objectives, 
which contradicts the fact that the emotional expression within a single utterance cannot be adequately conveyed with just one emotion label. 
Additionally, though some methods \cite{fayek2017evaluating,xia2021temporal,feng2022semi} attempted to incorporate local information by constructing emotional pseudo-labels, there is still room for improvement in both the quality of pseudo-labels and their recognition performance.

Hence in this study, we propose a novel HuBERT-based GMP-TL (Gender-augmented Multi-scale Pseudo-label Transfer Learning) framework for SER, as shown in Fig. 1. 
Overall, our primary emphasis is on two critical dimensions: \textbf{the meticulous acquisition of high-quality frame-level emotional pseudo-labels} and \textbf{the comprehensive utilization of both frame-level and utterance-level emotional labels}. 
The key insight behind is that the captured frame-level pseudo-labels of each speech signal could provide valuable complementary local details for modeling speech emotion.
To achieve this goal, we design three continuous stages in the proposed GMP-TL workflow.
First, with the guidance of utterance-level emotional labels, we train a pretrained-HuBERT based SER model implemented with multi-task learning (emotion and gender classification) and unsupervised multi-scale k-means clustering on different HuBERT layers to acquire the high-quality gender-augmented multi-scale pseudo-labels (GMPs). The training loss function is cross-entropy (CE).
\begin{figure*}[htbp]
\centering
	\includegraphics[height=9.6cm,width=!]{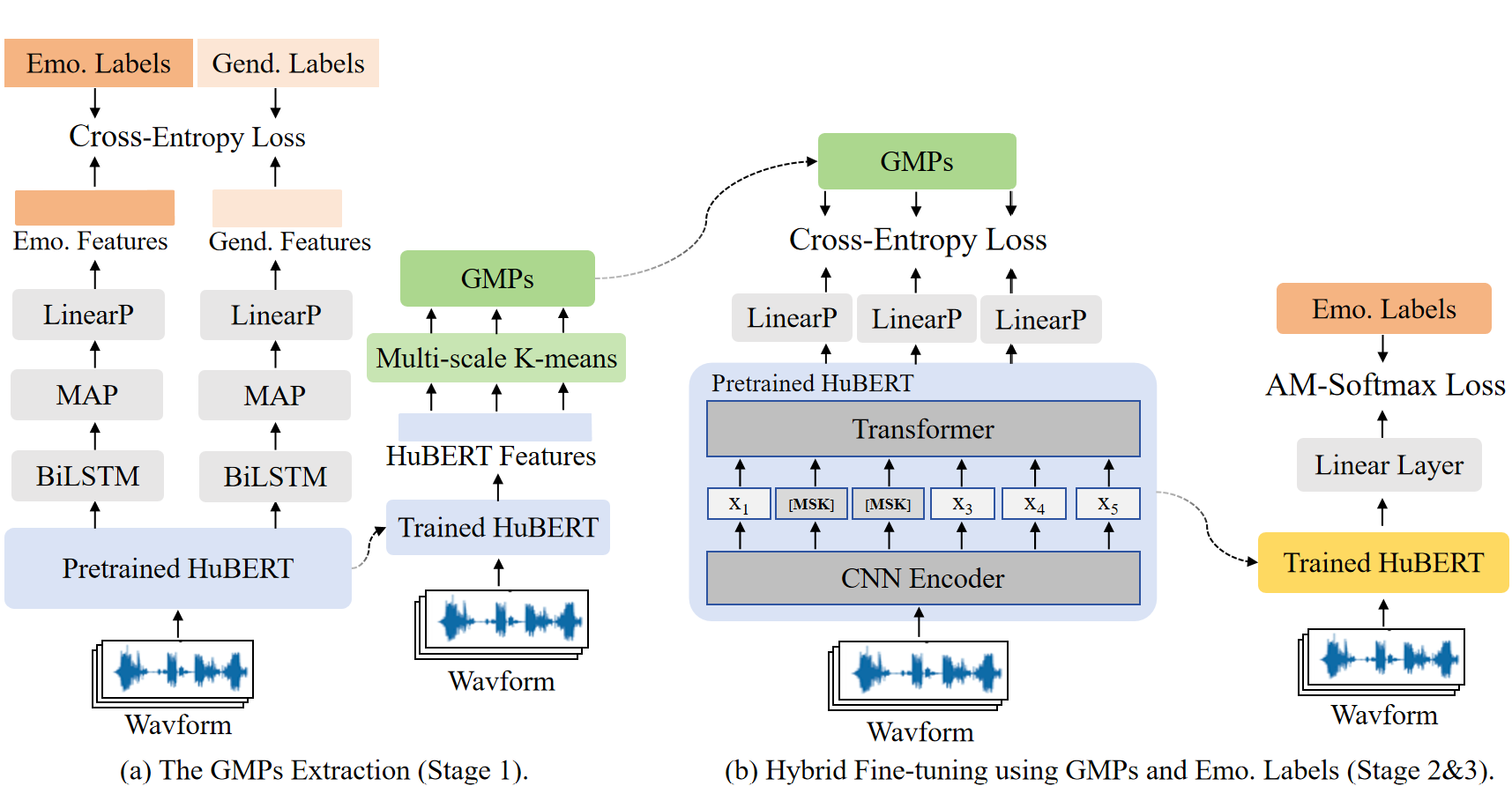}
	\caption{Overview of the proposed GMP-TL framework. BiLSTM is bi-directional LSTM, MAP indicates mean average pooling, LinearP represents linear projection module, Emo. and Gend. are the abbreviations of emotion and gender.}
\end{figure*}
Subsequently, to better leverage the frame-level GMPs and utterance-level emotion labels, we present an effective two-stage hybrid loss based fine-tuning (Hybrid-FT) strategy to further optimize the GMP-TL SER workflow.
To elaborate, we first employ the obtained GMPs to guide the pretrained-HuBERT based SER model using CE loss as well, in order to incorporate the beneficial information of GMPs for emotion identification. 
Afterwards, we utilize the AM-Softmax (AMS) loss \cite{wang2018additive} to fine-tune the proposed SER model based on the trained HuBERT of stage 2, under the supervision of
utterance-level emotion labels. 
In this way, our proposed GMP-TL framework excels in capturing fine-grained contextualized representations of speech emotion, leading to a substantial performance improvements.

% The remains of this paper are structured as outlined below: Section \ref{sec:Related Work} offers an overview of the related studies, while Section \ref{sec:METHODOLOGY} provides in-depth insights into the proposed GMP-TL SER framework. Experimental settings and results are detailed in Section \ref{sec:EXPERIMENTS}, and concluding remarks are drawn in Section \ref{sec:CONCLUSIONS}.

\section{Related Work}
\label{sec:Related Work}
In this section, we provide a brief overview of relevant works, including the HuBERT-based SER methods, attribute-based multi-task learning for SER, and pseudo-label based SER.

\subsection{HuBERT-based SER Models}

HuBERT stands out as a prominent pre-trained speech model renowned for its effectiveness in mastering speech representation learning. Adopting a self-supervised learning framework, HuBERT undergoes pre-training on extensive speech datasets and employs iterative unsupervised clustering to generate pseudo-labels for each training phase. This process enables the HuBERT model to learn efficient and robust feature representations.
Leveraging HuBERT's advanced capabilities in speech representation learning, numerous HuBERT-based SER methods have been proposed. 
For instance, \cite{fang23b_interspeech} presented a block architecture search strategy to explore downstream transfer of HuBERT-based features for emotion recognition. \cite{zhang23y_interspeech} introduced an optimal transport approach for cross-domain SER using the pre-trained HuBERT model.

\subsection{Attribute-based Multi-task Learning for SER}
Due to the multitude of attributes present in speech signals, numerous recent works \cite{li2019improved,ghriss2022sentiment,pan2023gemo,pan2023msac,zhang2022selective} aim to leverage these attributes to enhance SER. Among them, attribute-based multi-task learning emerges as an effective approach. It enhances the recognition performance of SER models by integrating auxiliary tasks, including gender classification, automatic speech recognition, and so forth. 
For instance, Li \emph{et al}. \cite{li2019improved} introduced a multitask learning (emotion and gender classification) based SER method and obtained great recognition results. 
Ghriss \emph{et al}. \cite{ghriss2022sentiment} advocated an end-to-end multi-task approach which employed a sentiment-aware automatic speech recognition pre-training for emotion recognition.
Zhang \emph{et al}. \cite{zhang2022selective} investigated an effective combination approaches on the basis of multi-task learning, with a focus on the style attribute.

\subsection{Pseudo-label based SER}
Constrained by the availability of labeled datasets, the majority of current methods normally rely on utterance-level emotion labels to achieve SER. 
Nonetheless, according to existing research \cite{fayek2017evaluating,xia2021temporal,feng2022semi}, utterance-level emotion labels of speech utterances may not be as accurate, and they suggest that introducing frame-level pseudo-labels can be beneficial in modeling speech emotion.
In \cite{xia2021temporal}, the authors presented Seg-FT, which investigated the significance of temporal context for SER and advocated a segment-based learning objective to leverage local features.
In \cite{fayek2017evaluating}, a dynamic frame based formulation was presented to achieve emotion recognition.
In \cite{feng2022semi}, the researchers designed Semi-FedSER, a semi-supervised federated learning approach for SER using multi-view pseudo-labeling.

\section{METHODOLOGY}
\label{sec:METHODOLOGY}
This section introduces the overall architecture of our proposed GMP-TL workflow, as depicted in Fig. 1. 
First, we provide a detailed exposition of the frame-level GMP extraction phase. 
Subsequently, the principles and specific implementation of the proposed Hybrid-FT approach are elucidated.

\subsection{Frame-level GMPs Extraction}
Intuitively, human speech is not likely to be consistently characterized by one single emotion, especially in a long utterance. 
Therefore, to bridge this gap, we believe that it is necessary to extract and incorporate high-quality frame-level pseudo-labels to train the SER model, whose overall schematic diagram is outlined in Fig. 1 (a).

Concretely, we initiate the process by employing the pre-trained HuBERT model as a feature extractor. The derived features are subsequently fed into a bi-directional LSTM network, and the impact of the temporal dimension is alleviated through an average pooling operation. Following this, the acquired features undergo a linear projection module that consists of two linear layers and one ReLU activation layer. 
Furthermore, drawing inspiration from relevant literature \cite{li2019improved,pan2023gemo,nediyanchath2020multi}, we incorporate a multi-task learning strategy into the training process to introduce gender information within speech signals. In this way, the overall model is capable of capturing more feasible emotional representations.
Ultimately, the entire framework is trained using the cross-entropy (CE) loss function, guided by utterance-level emotion and gender categorical labels.
Consequently, the final loss $L_{Total}$ of this stage is formulated as:
\begin{equation}
    \begin{split}
        L_{Total} = \alpha_{e} L_{Emo} + (1-\alpha_{e}) L_{Gender}
    \end{split}
\end{equation}
where $L_{Emo}$ is the CE-loss of emotion attribute, $L_{Gender}$ is the CE-loss of gender attribute, $\alpha_e$ is a parameter to adjust $L_{Emo}$ and $L_{Gender}$. In our case, $\alpha_e$ is set to 0.9.

Afterward, in order to generate frame-level GMPs of higher quality, we attempt to use features from different layers of HuBERT and perform multi-scale unsupervised k-means clustering, a departure from the conventional practice of using features from the final layer. In our case, the features from the third-to-last layer gains the best performance. The number of cluster centers is empirically set to 64, 512, and 4096, respectively.

\subsection{CE-loss based Fine-tuning using GMPs}
In the second stage, we adopt the CE-loss based model fine-tuning strategy that uses the obtained frame-level GMPs to retrain and optimize the HuBERT-based SER model. The concrete components of which are illustrated in the left side of the Fig. 1 (b).

As depicted in the figure, we adhere to the conventional masking operation of HuBERT and implement a linear projection module, which comprises two linear layers and one ReLU activation layer following the HuBERT model, to accurately harness and align the GMPs. The entire model is trained using CE loss as well.
In light of this means, the frame-level GMP-based fine-tuning method also aligns well with the original pre-training objectives of the HuBERT model, ensuring an effective utilization of information from both the original pre-trained HuBERT model and the obtained frame-level GMPs of previous phase.

\subsection{AMS-loss based Fine-tuning via Emotion Labels}
Due to many factors such as the inherent ambiguity in speech emotion, an issue arises where speech emotion categories show similarity between different classes and differences within the same classes. 

To alleviate this problem, in the last fine-tuning stage, we employ the AMS loss to fine-tune the HuBERT-based SER model under the guidance of utterance-level emotion labels. 
By incorporating a constraint vector margin, this approach is able to increase the inter-class distance between different emotion categories and reduce the intra-class distance within the same emotion category, thus enhancing the model's recognition performance.
Moreover, it is worth noting that we utilize the HuBERT model trained in the second stage as the feature extractor for this phase, with its comprehensive system layout presented in the right side of the Fig. 1 (b).

As a result, the final loss function for the fine-tuning stage can be defined as:
\begin{equation}
    \begin{split}
        L = \!-\frac{1}{N}\!\sum_{i=1}^N\!\log
        \frac{e^ {s(\cos(\theta_{y_i, i}) - m)}}
        {e^{s(\cos(\theta_{y_i, i}) - m)} + \sum_{j\neq{y_i}}e^{s(\cos(\theta_{j, i}))}}
    \end{split}
\end{equation}
\begin{equation}
    \begin{split}
        \cos(\theta_{j, i}) = \frac{{x}_{i}^{T}w_j}
        {\left\Vert x_i \right\Vert \left\Vert w_j \right\Vert}
    \end{split}
\end{equation}
\noindent
where $L$ represents the ultimate loss, $x_i$ and $y_i$ represent the feature vector and label of the $i$th sample, $w_j$ corresponds to the feature vector of class $j$, and $\theta_{j, i}$ signifies the angle between $x_i$ and $w_j$. N denotes the batch size, while s is the scaling factor, and m is the additive margin. In our specific scenario, the values for m and s are set at 0.2 and 30.

\section{EXPERIMENTS}
\label{sec:EXPERIMENTS}
In this section, we first outline the experimental setups of this study, encompassing the experimental database and implementation details. 
Second, a comprehensive comparison and analysis of the proposed approach with existing SOTA unimodal and multimodal SER methods is given.  
Finally, to examine the validity of our proposed GMP-TL framework, we conduct an ablation study.

\subsection{Experimental Setups}
\subsubsection{Experimental Datasets}

In this work, we conduct extensive experiments using the most challenging IEMOCAP \cite{IEMOCAP} corpus, which has been widely recognized for evaluating SER systems. IEMOCAP consists of scripted and improvised interactions recorded in five sessions with the participation of 10 actors (5 male and 5 female), and the dataset is labeled by three annotators. 

To facilitate a fair comparison with other SOTA methods, we merge "excited" and "happy" into the "happy" category, resulting in four emotion categories: angry, happy, neutral, and sad.

\subsubsection{Implementation Details}

In all experiments, we use Adam to optimize the proposed GMT-TL framework with an initial learning rate of 1e-4 and a batch size of 64. Besides, due to the limitations of computational resources, we employ the pre-trained HuBERT-base model, which is accessible on an open-source website$^1$.
{
\let\thefootnote\relax
\footnote{$^1$https://huggingface.co/facebook/hubert-base-ls960}
}

As for evaluating metrics, we adopt weighted average recall (WAR) and unweighted average recall (UAR) in the context of the speaker-independent setting.
Additionally, we perform a standard 5-fold cross-validation, aligning with current SOTA SER methods, to evaluate our GMT-TL. Within each fold, one session is reserved as the test set, and we calculate the average of the acquired UAR and WAR across all folds.

\subsection{Main Results}

To examine the effectiveness of the proposed GMP-TL method, we first compare its recognition performance with SOTA unimodal and multimodal SER approaches, with the results summarized in Table \ref{tab:comparison-results}.

\begin{table}[htbp]
    \centering
    \caption{Recognition comparison of SOTA SER methods on IEMOCAP. A and T are audio and text modalities, respectively.}
    \label{tab:comparison-results}
    \renewcommand{\arraystretch}{0.9} 
    \begin{tabular}{C{35mm}C{13mm}C{13mm}C{12mm}}
        \toprule
        Model                      &  Modalities  &   UAR   &  WAR   \\
        \midrule
        Seg-FT \cite{xia2021temporal}  &  A  &  66.9  &  65.4   \\
        Joint Network \cite{hu2023joint}  &  A  &  73.3  &  72.5   \\
        TAP  \cite{Gat}            &  A  &  74.2  &  -      \\
        P-TAPT \cite{Chen}         &  A  &  74.3  &  -      \\
        BAS \cite{fang23b_interspeech} &  A  &  75.0  &  74.3  \\
        UDA  \cite{Morais}         &  A  &  77.8  &  77.4  \\
        SA-CNN-BLSTM \cite{li2019improved}  &  A  &    82.8    &  81.6  \\
        \midrule
        UATMF \cite{10096586}  &  A + T  &  79.6  &  78.3  \\
        MMER \cite{ghosh2022mmer}  &  A + T  &    -    &  81.20  \\
        GEmo-CLAP \cite{pan2023gemo}  &  A + T  &   83.2    &  81.4  \\
        DC-BVM \cite{miao2024dc}  &  A + T  &   83.7    &  82.5  \\
        \midrule
        GMP-TL (Ours)   &  A  &  82.0  &  80.0  \\
        \bottomrule
    \end{tabular}
\end{table}

As illustrated in the above table, it is apparent that our GMP-TL achieves superior recognition results.
To be specific, when compared with SOTA unimodal SER methods including the Pseudo-label based Seg-FT model \cite{xia2021temporal}, the proposed GMP-TL attains the secondary best UAR and WAR of 82.0\% and 80.0\%. 
In addition, the proposed GMP-TL also demonstrates competitive recognition results as opposed to SOTA multimodal SER approaches, showcasing the effectiveness of our method.

\subsection{Ablation Study}
To comprehensively evaluate the validity of our design, ablation studies are performed. 

\subsubsection{Effect of key components in GMT-TL.}

We first validate the effectiveness of the two crucial components, i.e., GMPs and Hybrid-FT, of the proposed GMT-TL in Table \ref{tab:ablation-results1}.

\begin{table}[htbp]
    \centering
    \caption{The effect of key components on the performance of the GMP-TL. CE-FT is CE-loss based fine-tuning method, MPs denote multi-scale pesudo-labels.}
    \label{tab:ablation-results1}
    \renewcommand{\arraystretch}{0.9} 
    \begin{tabular}{C{7mm}C{9mm}C{11mm}C{17mm}C{9mm}C{10mm}}
        \toprule
        MPs &  GMPs &  CE-FT  & Hybrid-FT    &   UAR   &  WAR   \\
        \midrule
        \checkmark  &            &  \checkmark  &              &  78.0    &  75.8  \\
        \checkmark  &            &              &  \checkmark  &  79.8    &  77.6  \\
                    & \checkmark &  \checkmark  &              &  80.2    &  78.4  \\
                    & \checkmark &              &  \checkmark  & \textbf{82.0} &  \textbf{80.0}  \\
        \bottomrule
    \end{tabular}
\end{table}

From the table, we can easily observe that without the proposed frame-level GMPs, the performance of our GMP-TL framework has dropped by at least 2\% when just using the MPs, which proves our strategy that incorporating GMPs into the speech emotion modeling are beneficial.
Additionally, without the Hybrid-FT strategy, the performance of GMP-TL drops as well compared to using CE-FT, indicating the effectiveness of the proposed Hybrid-FT approach.

\subsubsection{Effect of GMPs Using Different Feature Layers.}

Moreover, we study the effect of GMPs based on different feature layers of HuBERT within the proposed GMP-TL framework. Results are shown in Table \ref{tab:ablation-results2}.

\begin{table}[htbp]
    \centering
    \caption{The effect of GMPs based on different HuBERT's feature layers on the performance of the GMP-TL workflow.}
    \label{tab:ablation-results2}
    \renewcommand{\arraystretch}{0.8} 
    \begin{tabular}{C{25mm}C{18mm}C{15mm}C{15mm}}
        \toprule
        Method                   &  Layer ID   &   UAR   &  WAR   \\
        \midrule
        \multirow{6}{*}{GMT-TL} &  -1   &  80.4    &  79.0  \\
                                 &  -2   &  80.2    &  \textbf{\emph{79.7}}  \\
                                 &  -3   &  \textbf{82.0}    &  \textbf{80.0}  \\
                                 &  -4   &  \textbf{\emph{81.9}}    &  \textbf{\emph{79.7}}  \\
                                 &  -5   &  80.6    &  78.6  \\
                                 &  -6   &  80.6    &  78.1  \\
        \bottomrule
    \end{tabular}
\end{table}

It is evident that under the proposed GMP-TL workflow, the performance of GMPs based on the final layer features of HuBERT is suboptimal. Conversely, GMPs relying on the features from the third-to-last and fourth-to-last layers demonstrate superior results, which may be attributed to the final layer features of the HuBERT model containing more semantic information rather than emotion-related details.

\section{CONCLUSIONS}
\label{sec:CONCLUSIONS}
In this work, we propose GMP-TL, an effective GMPs-based transfer learning framework for SER. Initially, GMP-TL employs a pre-trained HuBERT model, implemented with multi-task learning and multi-scale k-means clustering strategies, to obtain high-quality frame-level GMPs. 
Moreover, an efficient two-stage hybrid loss based fine-tuning approach is introduced to further optimize the GMP-TL by comprehensively leveraging frame-level GMPs and utterance-level emotion labels.
Experiments on IEMOCAP show that our GMP-TL not only attains superior performance compared to SOTA unimodal SER methods but achieves competitive results compared to multimodal SER approaches, demonstrating the effectiveness of the proposed approach.
In future work, we aim to develop more accurate frame-level emotional pseudo-labels by integrating additional relevant speech attributes, thereby further enhancing the proposed GMP-TL framework.

\vfill\pagebreak
\label{sec:refs}

\footnotesize
\bibliographystyle{ieee}
\bibliography{GMP_TL}

\end{document}